\documentclass[letterpaper, 10 pt, conference]{ieeeconf}  

\usepackage{graphicx}      

\newcommand{\barr}[2]{\begin{array}{#1}#2\end{array}}
\newcommand{\beq}{\begin{equation}}
\newcommand{\eeq}{\end{equation}}

\newcommand{\R}{\rm{I\kern-2pt R}}

\IEEEoverridecommandlockouts                              

\title{\LARGE \bf
Proprioceptive feedback paradigm for safe and resilient motion control
}

\author{Mrdjan Jankovic$^{1}$
\thanks{$^{1}$Southwest Research Institute,
        Ann Arbor, MI 48108, USA
        {\tt\small mrdjan.jankovic@swri.org}}%
}

\begin{document}

\maketitle
\thispagestyle{empty}
\pagestyle{empty}

\begin{abstract}
Proprioception is a human sense that provides feedback from muscles and joints about body position and motion. This key capability keeps us upright, moving, and responding quickly to slips or stumbles. In this paper we discuss a proprioception-like feature (machine proprioceptive feedback -- MPF) for motion control systems. An unexpected response of one actuator, or one agent in a multi-agent system, is compensated by other actuators/agents through fast feedback loops that react only to the unexpected portion. The paper appropriates the predictor-corrector mechanism of decentralized, multi-agent controllers as ``proprioceptive feedback" for centrally controlled ones. It analyzes a nature and degree of impairment that can be managed and offers two options  -- full-MPF and split-MPF -- with different wiring architectures as well as different stability and safety properties.  Multi-vehicle interchange lane-swap traffic simulations confirm the analytical results. 
\end{abstract}

\section{INTRODUCTION}
A key requirement of human motion control is to maintain safety while walking, running, lifting objects, etc. Human motion may be subject to a sudden interruption -- a slip or stumble -- that requires a very fast response to prevent a fall.  An alternative would be a more conservative motion such as ASIMO robot's stable walk \cite{asimo}, which is less efficient and less responsive than human slightly-unstable gait. An analogous situation appears in standard robust control: preparing for the worst case disturbance, the system operates conservatively. 
 In most cases in the control literature, the goal  of robust control is to resist (unintended) change, maintain its original operating regime, and function under a wide range of conditions. The ultimate goal is {\em stable operation} in the presence of changing parameters or persistent disturbances. 

Recent  multi-agent system work leverages {\em instability} to provide liveness \cite{jankovicTCST, jankovicAR}, agility \cite{jankovic_ACC26}, and efficiency \cite{deshpande_ACC26}.
These papers use Control Barrier Functions (CBFs) for safety guarantees. The mechanism that produces instability was originally introduced to handle  unexpected behavior  due to agents' incomplete and asymmetric information. It has turned out that it does more, providing {\em resilience}\footnote{Resilience definition: ``ability of a system to absorb sudden disruption, recover quickly, and adapt or bounce back after a disturbance has occurred."} thanks to  very fast predictor-corrector feedback loops.  

\begin{figure} 
	\centering
	\includegraphics[width=74mm]{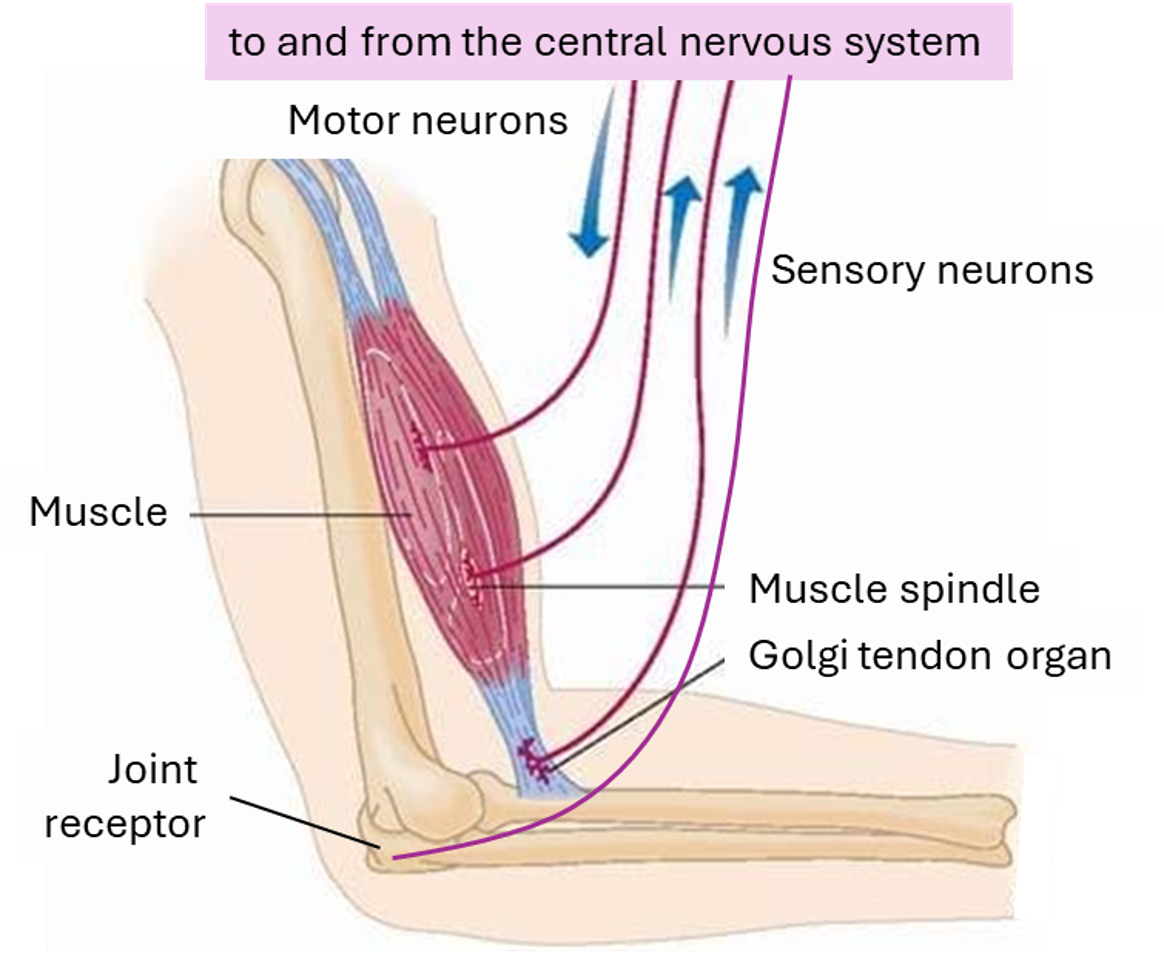} \hfill
    
   \vspace{-3mm} \hfill (a) \ \ \  

    \includegraphics[width=85mm]{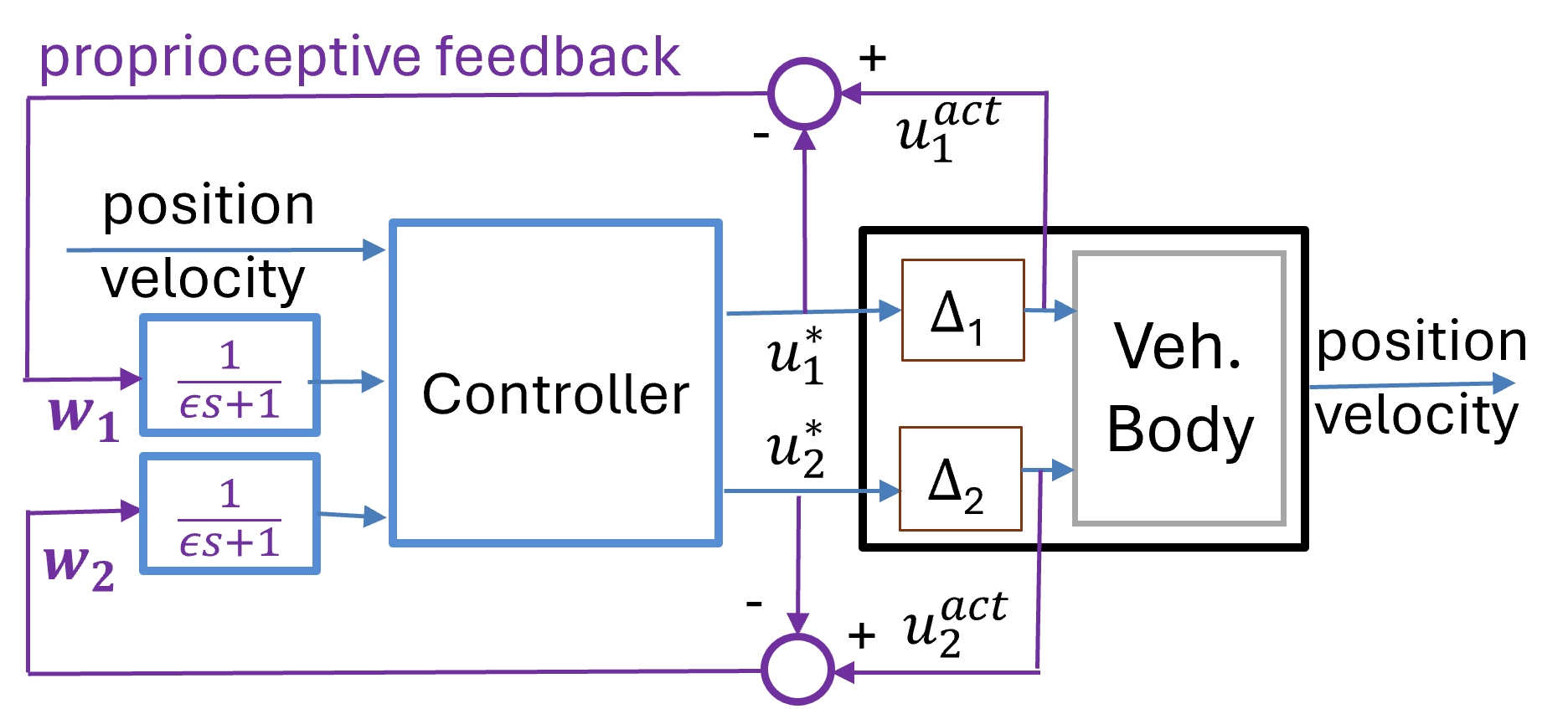}
    
   \vspace{-3mm} \hfill (b) \ \ \ 
       \vspace{-2mm}
	\caption{a) Proprioceptive feedback structure in human muscles; b) machine proprioceptive loops for a two degree of freedom controller -- e.g., steering ($u_1$) and acceleration ($u_2$) in a vehicle -- considered in this paper.}
	\label{fig:muscle_loops}
    \vspace{-4mm}
\end{figure}

A similar mechanism  exists for postural control in humans and animals: the proprioceptive feedback (\cite{proske, tuthill, hanley}). Proprioception is a ``muscle sense'' that enables perception of the position and motion of the body in space independent from vision, vestibular, or other senses. Proprioceptive feedback connects motor neurons with special sensory neurons in muscles, tendons and joints as illustrated in Fig. \ref{fig:muscle_loops}.a.  An illustrative proprioceptive feedback use case is an instinctive reaction to tripping. In response to the leg motion being interrupted, a hand immediately reaches forward and the other leg extends to prevent a fall, before the person actually starts to fall. As discussed below in Section \ref{sec:Proprioception}, there is a part of human proprioception-related response that does not engage the brain at all -- the ``spinal reflex'' which is much faster than a vision based one. 

The goal of this paper is  to transfer the proprioceptive feedback paradigm to design of motion control systems. The motivation includes
\begin{itemize}
    \item Establishing a biological counterpart for a control algorithm may provide insights that lead to improvements (e.g., the new split-MPF algorithm in this paper);
    \item Connecting an algorithm with a biological feature honed by evolution over hundreds  of millions of years (proprioception is present in insects \cite{tuthill});
    \item Explaining observed interchange traffic results;
    \item Providing control theoretic insights may help understanding human/animal proprioceptive feedback. 
\end{itemize}

 We consider nonlinear systems with safety constraints that must hold at all times, including during a sudden adverse event.  
A block diagram of a control system augmented with the Machine Proprioceptive Feedback (MFP) loops is shown in Fig \ref{fig:muscle_loops}.b. The controller computes the intended control actions and monitors their execution. $\Delta_1$ and $\Delta_2$ represent uncertain static or dynamic blocks  assumed in the design phase to be equal to 1.
The difference is fed back into the controller to compensate for any discrepancy. The first order filters in Fig. \ref{fig:muscle_loops}.b break potential algebraic loops. The time constant, denoted by $\epsilon$ to indicate a small quantity, is selected such that the filters are much faster than the system dynamics and the safety-related CBF dynamics.  

It is not easy to assure stability of very fast feedback loops in the presence of uncertain, potentially slower dynamics as is the case here. Stability of feedback loops with unknown dynamics has been studied in the control literature  (e.g.,  Absolute Stability, Chapter 7 in \cite{khalil}; Integral Quadratic Constraints (IQC) \cite{megretski}).  Within safety context, CBF safety filters with IQC for robustness to input uncertainty has been considered in \cite{seiler}. Without the proprioceptive $w$-loops, IQC tends to produce a conservative controller. 

A contribution of this paper is showing that, for the ``full-MPF"  depicted in Fig \ref{fig:muscle_loops}.b, the fast-loop stability and safety are assured if a connection of $\Delta_1$ and  $\Delta_2$ is input strictly passive (ISP), see Chapter 6 in \cite{khalil}. The ISP is related to inputs and outputs being ``in-phase" and having a direct feed-through.  The class of ISP systems includes the case when one of the control channels suddenly stops working. Through the fast MPF-loops, the controller  very quickly transfers the safety responsibility to the other available actuator. 

If ISP cannot be guaranteed a-priori (e.g., even a short delay is not ISP, not even passive), the ``split-MPF" option is introduced. This new architecture has a sub-controller responsible for only one actuator while seeing the other's deficiency (like the other limb in the  cross-extensor spinal-reflex). 
The split-MPF controller has much broader fast-loop stability, including delays, and works well if  one actuator is impacted. It does not, however,  guarantee CBF constraint satisfaction if both actuators are impaired simultaneously. 

We illustrate the results with  highway interchange lane-swap traffic  for connected and automated vehicles (CAV) \cite{jankovic_ACC26}. In this complex multi-agent system, there are four actuators  for every vehicle-to-vehicle collision constraint, while each vehicle may have more than one such constraint. Unexpected actuator impairments include random vehicles with  sudden loss of acceleration or steering ability, and/or filtered/attenuated actuator response.  Extensive Monte Carlo (MC) simulations confirm the analytical results.

\section{Proprioceptive feedback}
\label{sec:Proprioception}
Proprioception is the human's ``sixth sense" that provides feedback about body's position in space as well as the level of muscle exertion. It is largely subconscious, but gives us the feeling of being ``in the body." A similar mechanism exists in animals including insects. In this section we briefly review proprioception and proprioceptive feedback based on \cite{proske, tuthill, hanley}. We concentrate on resilience and recovery from a disruption, while noting that  proprioception also provides means to coordinate limbs to produce movement. Compared to other human sensory feedback systems, of specific interest here is that  proprioception is distributed, assigned to a specific actuator, and fast. The concept is used to motivate our work on resilient multi-agent systems and our MPF design choices. 

Proprioceptive sense is based on mechanoreceptors in our muscles, tendons, and joints, Fig \ref{fig:muscle_loops}.a. Motor neurons activate the muscle to contract, while the sensory neurons send back the information about the amount of contraction (length), velocity, as well as the level of force applied. Similarly, the Golgi tendon organ provides  the load or tension in the tendon while the joint receptor typically activates at the physical limits of the joint motion. Together, these sensory receptors provide feedback to the spinal cord and the cerebellum that is not directly related to the ultimate objective, say, a hand reaching an object. Rather it shows if the local actuator (the muscle) is executing its assigned task as expected, similar to the $w$-feedback in Fig \ref{fig:muscle_loops}.b. 

\begin{figure}
 \includegraphics[width=85mm]{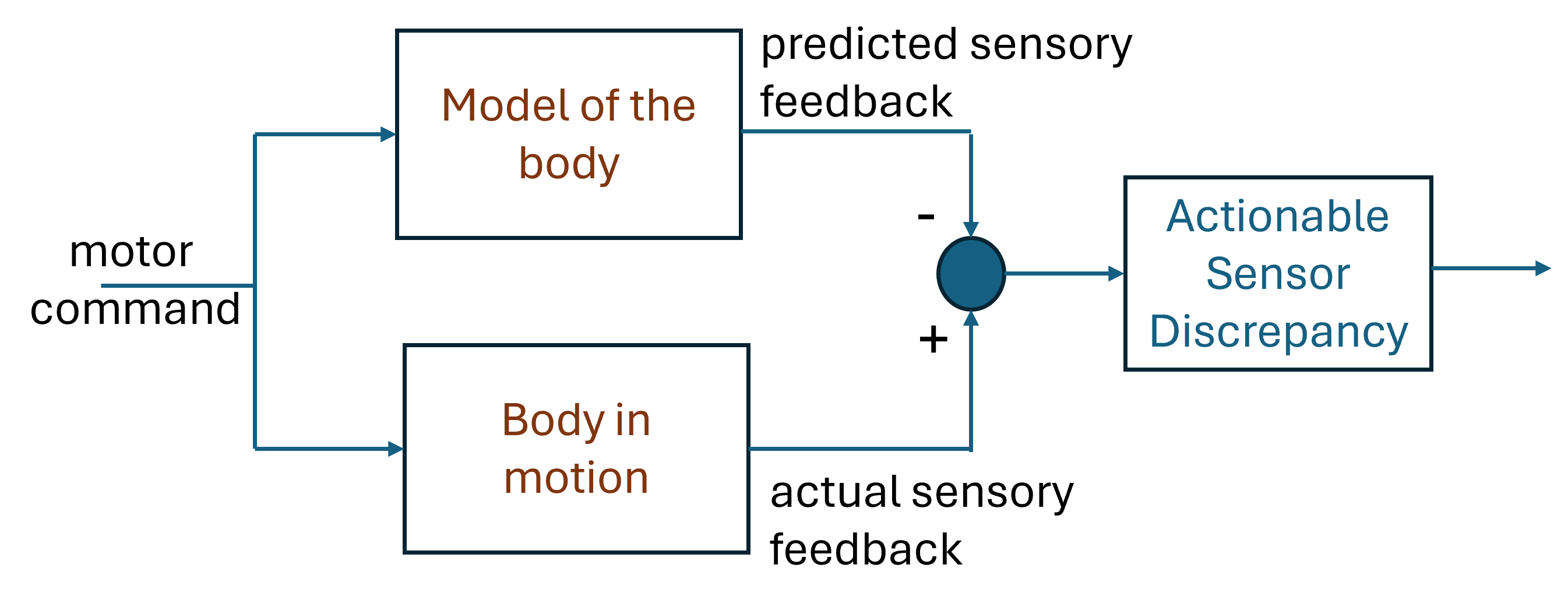}   
      \vspace{-3mm}
   \caption{A model of a ``difference calculator" that produces an actionable signal from the proprioceptive feedback (based on \cite{proske, bays}).}
	\label{fig:difference}
   \vspace{-3mm}
\end{figure}

It is generally understood  that the proprioceptive feedback operates on a difference between what is expected and what actually occurred \cite{proske, bays}. In human motion control, there are situations when unexpected increase in resistance to muscle motion results in increased effort for the same muscle group (e.g. lifting a weight heavier than expected).  It is assumed that this reaction is conscious (i.e. not reflexive), but only the difference between the expected and actual motion is perceived as shown in Fig. \ref{fig:difference}.

One interesting human reaction related to our MPF work is spinal reflex, in particular ``cross-extensor" reflex, see Section 28 in \cite{hanley}. As the name suggests, the reaction does not include the  brain at all and operates within the muscle groups, motor and sensory neurons, and the spinal cord. Thus, the (re)action is very fast: tens rather than hundreds of milliseconds. If a person steps on a sharp object, the spinal reflex withdraws (contracts muscles of) the affected leg. The cross-extensor reflex extends the other leg automatically to maintain balance and prevent a fall. 

 As mentioned in the Introduction, human gait is unstable. Stopping a leg mid-movement (tripping) can cause a fall. When the stopped movement is detected, a  more complex response than the spinal reflex may be initiated. It  involves muscle groups from the torso and other limbs (e.g. reaching for the handrails) with  unconscious brain coordination. In this case, as well as in the case of the cross-extensor spinal reflex, the perceived deficiency of one actuator (stopped muscle motion in a limb) does not produce an effort to actuate the same group more. Instead, it tries to recover by fast reactions of other actuators. This is a good analogy of how the split-MPF works.  In summary, properties of the proprioceptive feedback that are relevant for MPF design are
\begin{itemize}
    \item The proprioceptive sense is distributed and collocated with the actuators (muscles).
    \item Proprioceptive feedback is safety critical and must be fast. 
    \item The reaction to unexpected motion  of a muscle group may be to further actuate the same group (probably not a reflexive reaction) or deactivate the affected group and compensate with other muscle groups (likely an unconscious, reflexive reaction).  
\end{itemize}

\section{Safety Critical Systems}
\label{CBF}
One key ingredient  for the concept of machine proprioceptive feedback is  existence of a safety constraint that the system  must satisfy at all times. One example is road vehicles that must maintain a predetermined minimum distance. Here, we employ Control Barrier Functions \cite{ames, ames_IEEE17, ames_ECC19} to enforce safety. Following \cite{jankovicRCBF}, we consider input-affine
nonlinear systems with disturbance: 
\begin{equation} \dot x = f(x) + g(x) u + p(x) w \label{dynamics} 
    \end{equation}
The vector $x\in \R^n$ is the system's state, $u\in {\mathcal U} \subseteq \R^{n_u}$ is the control input, and $w\in {\mathcal W} \subseteq \R^{n_w}$ is an external disturbance. Here, the disturbance input relates to actuator unexpected behavior as shown in Figs. \ref{fig:muscle_loops}.b and Fig. \ref{fig:difference}, the difference of what is commanded (expected) and what is observed.

The system has an admissible (safe) set ${\mathcal C} = \{ x\in {\R^n}: h(x) \ge 0\}$ defined by a differentiable function $h(x)$ (e.g., vehicle-to-vehicle distance). The following definition combines exponential \cite{nguyen} and robust \cite{jankovicRCBF} CBFs. \vspace{2mm}  \\
{\bf Definition 1}: The function $h(x)$ is an (exponential) Robust Control Barrier Function (RCBF) if there exists a constant $\lambda > 0$ such that 
\begin{equation}
   \max_{u\in {\mathcal U}}\min_{w\in {\mathcal W}}\{ \dot h(x,u,w) + \lambda h(x)\} >0    \label{ERCBF}
\end{equation}
The time derivative of $h$ is given by $\dot h = L_f h + L_g h\ u + L_p h\ w$, where $L_f h$ denotes $\frac{\partial h}{\partial x}f$. The safety constraint to be satisfied at all times is 
\beq L_f h + L_g h\ u + L_p h\ w +  \lambda h \ge 0 \label{safety} \eeq
 As long as the control satisfying (\ref{safety}) is used, $\dot h + \lambda h \ge 0$ and $h(t) \ge 0, \forall t > t_0$ provided that $x(t_0)$ is in the admissible set $\mathcal C$. Thus,  a system that starts in the safe set stays in the safe set.

Typically, CBF is used as a safety filter to override a baseline, or performance, control $u_0$ using a Quadratic Program (QP). In the case of human walking, $u_0$ could be a motor command that generates repetitive leg  motion. In the case of a vehicle, $u_0$ could represent steering and acceleration needed to follow the road at the  desired speed. The MPF QP takes the form
\begin{equation}
\barr{l}{ u^* = {\rm argmin}_{u \in {\mathcal{U}}}  \|u - u_0\|^2 \ \ {\rm such \ that} \\*[2mm]
L_f  h + L_g   h\ (u+ w) + \lambda h = 
a + b (u + w ) \ge  0  } \label{EPF}
\end{equation} 
with $a =  L_f h(x) + \lambda h(x)$, $b = L_g h(x) $, and $w$ assumed known. By definition of the RCBF, we know that the solution exists, that is, the QP is feasible.
In the next two sections, we assume that ${\mathcal U} = \R^{n_u}$, i.e. the control is unconstrained.

In the presence of input uncertainty $\Delta$, but without the $w$ term in (\ref{EPF}), the CBF constraint  becomes
\[ \dot h + \lambda h + b (\Delta(s)-I) u^* \ge 0\]
For simplicity, we assume that $\Delta = diag\{\Delta_1, \ldots, \Delta_{n_u}\}$ is a linear system with unknown gains and unknown dynamics in its $n_u$ channels. $(\Delta(s) - I)\varphi(t)$ notation is to be understood as the linear operator acting on the time varying input $\varphi(t)$. The quantity $b(\Delta(s)-I) u^*$ is not necessarily positive or small. Thus, $h(x)$ may go negative and the state $x$ may leave the safe set $\mathcal C$.

\section{Full Machine Proprioceptive Feedback}
\label{sec:singular_p}

 In the original decentralized predictor-corrector applications \cite{santillo, jankovicTCST}, the fast ``disturbance w" feedback was used to reconcile  differences between (the ego vehicle) local  copies and actual actions of other agents. The differences arose from incomplete and asymmetric information about $u_0$. In the MPF case, we deal with a problem of unexpected actuator response that brings up fast-loop stability issues.  
 
 As illustrated in Fig. \ref{fig:muscle_loops}.b, the  vector $w$ is the difference between the expected and the actual control action filtered by a very fast filter
 \beq \epsilon \dot w = -w + (u^{act} - u^*) \label{w-loop} \eeq
with the time constant $\epsilon  \ll  1/\lambda$. That is,  we tune the $w$-feedback  to be much faster than the CBF constraint dynamics and the system (\ref{dynamics}). Because the complete $w$-vector is used in (\ref{EPF}), we refer to this algorithm as {\em full-MPF}. 

The control input computed  from (\ref{EPF}) is
\beq u^* = u_0 - \frac{\min\{0,\ a + bu_0 + b w\} }{\|b\|^2} b^T \label{EPF_control} \eeq
Note that by the definition of RCBF, $b \not = 0$ when the constraint is active, i.e., when $a + bu_0 + b w \le 0$.  

We now consider $\eta_1 = \frac{b(x)}{\|b(x)\|}w$, a product of the slow and fast variables.  If the CBF constraint is active, using (\ref{w-loop}) we obtain
\beq \epsilon \dot\eta_1 = \frac{a+ b u_{act}}{\|b\|} + \epsilon  \left(\frac{d}{dt} \frac{b}{\|b\|} \right) w \label{sing_pert} \eeq
Note that $u_{act}$ depends on $\eta_1$ through  $\Delta$. This equation is in the standard singular perturbation form (see, \cite{khalil}, Chapter 11).  If this coupled system is asymptotically stable, 
the singular perturbation theory (Tikhonov's Theorem 11.1, \cite{khalil}) applies: for a positive $\delta$ arbitrarily  small, and $T$ arbitrarily large, if the time constant $\epsilon$ is sufficiently small, $a + bu_{act} = {\mathcal O}(\epsilon)$\footnote{In the standard Singular Perturbation notation, ${\mathcal O}(\epsilon)$ denotes a quantity that converges to 0 as fast as $\epsilon$} on an interval $[t_0+\delta, T]$,
with $t_0$ the initial time.  The CBF dynamics is $\dot h +\lambda h = a + bu^{act}$ and the singular perturbation result gives an estimate
\beq h(t) = h(t_0)  e^{-\lambda (t-t_0)} + {\mathcal O}(\epsilon) > {\mathcal O}(\epsilon) \eeq
for all $t \in [0,T]$. Thus, if the  $w$-loop is asymptotically stable (and remains fast), the CBF $h$ is always larger than a very small possibly negative quantity. To handle ${\mathcal O}(\epsilon)$, we need a small barrier margin. Typically, the safe set is made a little more conservative (i.e smaller), relying on 
CBF robustness to small disturbances, see \cite{xu_IFAC}. 

Similarly, if the CBF constraint is inactive, $a + b u_0 + b w \ge 0$, the control
is $u^* = u_0$ and, using Singular Perturbation, $w = u^{act} - u_0 + {\mathcal O}(\epsilon)$. Substituting into the constraint produces $\dot h + \lambda h = a + b u^{act} \ge {\mathcal O}(\epsilon)$. In this case, $u^*$ and $u^{act}$ are independent of $w$ and (\ref{w-loop}) is asymptotically stable.

\subsection{Input strict passivity and full-MPF loop stability}
\label{sec:passivity}

The key question  we need to answer is under what conditions  is (\ref{sing_pert}) asymptotically stable. For this purpose, we employ input strict passivity (see Chapter 6 of \cite{khalil} and Chapter 2 of \cite{sepulchre}). The notion of passivity originates from circuit theory where it was observed that passive RLC circuits have the relative 
phase between inputs (voltages) and outputs (currents) always within $\pm 90^\circ$. The negative feedback connection between two such systems is stable as well as passive itself.  

Because  fast asymptotic stability is needed, we use a stronger passivity property. A dynamical system with the state $z$, input $u$, and output $y$ is input strictly passive (ISP) if there is a positive definite ``storage function" $S$ such that 
\beq \dot S(z) \le y^T u - \nu u^T u \label{storage} \eeq
For exposition efficiency, we consider linear ISP systems
\beq \barr{l}{\dot z= Az + Bu\\
y = Cz + Du } \label{z_system} \eeq
with $A$ a Hurwitz matrix (i.e., $\dot z = A z$ is asymptotically stable). The ISP implies that the matrix $D$ is positive definite with the minimal eigenvalue $\lambda_{\min}(D) \ge \nu$. 
The system $\bar\Delta$ obtained by pre- and post-multiplying  $\Delta$ by a vector and its transpose,
$\bar \Delta =\frac{b}{\|b\|}\Delta \frac{b^T}{\|b\|}$, is also input strictly passive (\cite{sepulchre}, Proposition 2.11) with the same storage function $S$ and, in this case, the same supply rate $y^T u - \nu u^T u$. It doesn't matter if the vector $b$ is time varying or not. In other words, we do not assume that $\Delta$-dynamics  is $\frac{1}{\epsilon}$-fast.

Combining  (\ref{sing_pert}) and (\ref{z_system}) we obtain
\beq \barr{l}{\epsilon \dot \eta = -\frac{b}{\|b\|} D \frac{b^T}{\|b\|} \eta  + \frac{b}{\|b\|}C z + \psi_1(x) + {\mathcal O}(\epsilon) \\*[2mm]
\dot z= Az + B(-\frac{b^T}{\|b\|} \eta +\psi_2(x)) }\label{passive_loop} \eeq
where the terms $\psi_i(x)$ are easy to derive from  the control input $u^*$. The $(\eta,z)$ loop is a negative feedback connection of two passive systems and, thus, stable. Asymptotic stability follows from ISP and $A$ being Hurwitz.  Repeating the Singular Perturbation argument, we set $\epsilon \dot \eta = 0$ and, from (\ref{sing_pert}), $a + bu^{act} = {\mathcal O}(\epsilon)$ for $t\in [t_0 + \delta, \ T]$.  This, in turn,  assures that $h(t) > {\mathcal O}(\epsilon)$ 
for $t\in [t_0, \ T]$ as shown above. 

By restricting the uncertainty $\Delta$ to be ISP, we assure stability of the feedback loop and the $x$-system stays in the safe set. 
Passive systems include stand-alone gains $\kappa \in [0, \infty)$, first order filters with any positive gain, etc.  Strict input passivity, in addition,  requires a direct feed-through, $D > 0$.  We only need one of the subsystems $\Delta_i$ to be ISP (while others are just passive), provided that the corresponding $i$-th entry of the vector $b$ is bounded away from zero. 

The class of passive systems does not include second order filters or time delays. To illustrate a contrast with the split-MPF discussed next, we look at the case of equal delay in all channels: $\Delta(s) = e^{-\tau_d s}I$. 
Assuming that the delay is short enough so that the slow $b(x)$ stays approximately constant, from (\ref{sing_pert}) we obtain
\beq \epsilon \dot \eta = -\eta(t-\tau_d)  + \psi_3(x) + {\mathcal O}(\epsilon) \label{delay} \eeq
This system is unstable if $\tau_d > \frac{\pi}{2} \epsilon$. Moreover, to prevent undesirable under-damped response, we need $\tau_d < \epsilon$. Thus, the delay that could be tolerated is smaller than $\epsilon$, the  time constant of the fast system. 

\section{Split-MPF}
\label{sec:split_MPF}

Full-MPF replaced $b(\Delta(s) - I)u^*$ with likely much smaller ${\mathcal O}(\epsilon)$ error in the CBF constraint dynamics. In the process, it has acquired sensitivity to small delays and other non-passive dynamics. If the dynamics $\Delta$ are known, applying it to $u^*$  would approximately zero-out the feed into the $w$-filters (see Fig. \ref{fig:muscle_loops}). This approach  mimics the proprioceptive structure in Fig. \ref{fig:difference} as well as the internal feedback loops in human brain described in \cite{li_PNAS}. The latter paper explicitly credits these loops with compensating for delays. However, modeling $\Delta$ is often not an option because it is not known a-priori or it may appear as a result of a sudden event as illustrated by the interchange traffic example below.

\begin{figure}
 \includegraphics[width=85mm]{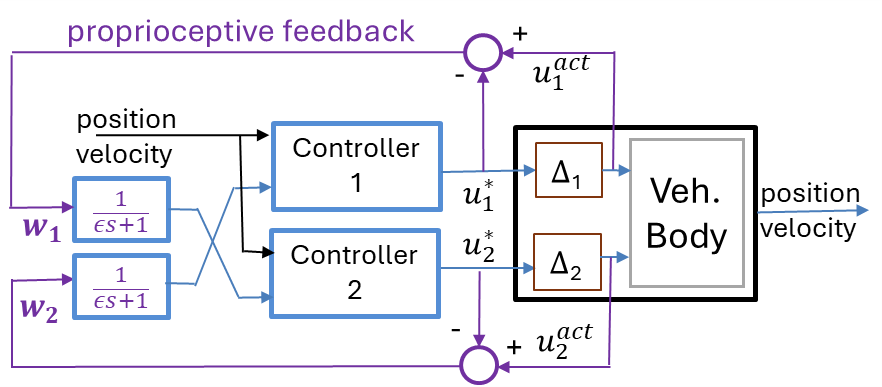}   
      \vspace{-2mm}
   \caption{Split-MPF configuration: separate controllers control each actuator while reacting to other actuator's unexpected response.}
	\label{fig:split-mpf}
   \vspace{-3mm}
\end{figure}

Here,  we propose an MPF mechanism that resembles cross-extensor spinal reflex (see the bullet points in Section \ref{sec:Proprioception}). While the full-MPF (partially) ``doubles down" on using the impaired actuator, the split-MPF  approach substitutes the missing contribution of the affected actuator with enhanced action of other ones as shown in Fig. \ref{fig:split-mpf}. In response, the affected actuator pulls back because it does not see its own impairment, but it does see the enhanced safety action of the other actuator(s). 

In the split-MPF case, each actuator has its own sub-controller. The approach is decentralized because the sub-controllers are using different information about actuator impairments.
For the actuator $i$, the QP controller is 
\begin{equation}
\barr{l}{ u_i^* = {\rm argmin}_{u_i \in {\mathcal{U}}}  \|u_i - u_0\|^2 \ \ {\rm such \ that} \\*[2mm]
a + b (u_i + w_i ) \ge  0  \\*[2mm]
\epsilon \dot w_i = -w_i + u^{act} - u_i^*, \ {\rm with} \  w_{ii} \equiv 0 } \label{S-MPF}
\end{equation} 
where $u_i = [u_{i1}, \ldots, u_{i n_u}]^T$ and $w_i = [w_{i1}, \ldots, w_{i n_u}]^T$. Except for $u_{ii}$, all other computed control inputs are local copies only used to compute $w_i$. The actual implemented control is the ``ego" output from every sub-controller: $u^* = [u_{11}^*, \ldots, u_{n_u n_u}^*]^T$. As before, $u^{act} = \Delta(s) u^*$.

To describe Split-MPF operation, let us just consider the 2-actuator system as shown in Fig. \ref{fig:split-mpf}. If the controls are unaffected by $\Delta$, $w$'s are equal to 0 and they both compute the same control input, $u_1^* = u_2^*$. If one of them, say $u_2$ is impacted by $\Delta_2$, $w_{12}$ inside Controller 1 becomes non-zero. As before, the singular perturbation provides $w_{12} =  u^{act}_2 - u_{12} + {\mathcal O}(\epsilon) $. The constraint in (\ref{S-MPF}) becomes
\[a + b_1 u_{11} + b_2 u_{12} + b_2 w_{12} = a + b_1 u_1^{act} + b_2 u_2^{act} +  {\mathcal O}(\epsilon)   \]
Thus, if the actuator 1 is unaffected, and the $w$-feedback is asymptotically stable and fast,  the safety of the system is assured within $ {\mathcal O}(\epsilon)$. 

This consideration reveals a drawback of the approach. If both actuators are impacted at the same time, the cross compensation may not be enough to keep the  error within $ {\mathcal O}(\epsilon)$. The problem arises only if both actuator $\Delta_i$s attenuate the control input. 
There is no problem if one actuator uncertainty amplifies the input. 

The advantage of this approach is that the stability of the $w$-loop is much broader than for full-MPF. If we reconsider the case from Subsection \ref{sec:passivity} where equal delay $\tau_d$ in both channels is considered, we obtain the fast dynamics as
\beq \epsilon \dot \eta = -\frac{2b_1^2 b_2^2}{\|b\|^4}(\eta + \eta(t-\tau_d) ) + \psi_4(x) + {\mathcal O}(\epsilon) \label{delay2} \eeq
In contrast to (\ref{delay}), here we have the direct feed-trough term as if the delay is an ISP system. While in the full-MPF case the delay must be limited to $\tau_d < \frac{\pi}{2}\epsilon$ for stability,  the system (\ref{delay2})
is stable for all $\tau_d$. This can be shown using Razumikhin stability theory (Theorem 4.3 in \cite{hale}). Obviously, with a very long delay in both channels, the safety cannot be guaranteed. Thus, split-MPF disconnects stability of the fast $w$-loop and CBF safety while for full-MPF asymptotic stability is more difficult to attain, but it guarantees safety.

\section{ Highway Interchange for CAVs}
\label{sec:Results}
The results of Section \ref{sec:singular_p} and \ref{sec:split_MPF} are illustrated on the interchange lane-swap traffic scenario  with the algorithms running the same ``IDA-fast" tuning described in \cite{jankovic_ACC26}. The main difference from \cite{jankovic_ACC26} is that the controllers have full information about everyone's intent\footnote{The MPF controllers would work with incomplete information, but our baseline no-MPF controller needs to know other agents' intent.}. The controller could be executed in a central roadside unit or distributed in the vehicles. The additional intent information could be provided by adding it to the standard vehicle-to-vehicle (V2V) Basic Safety Message (BSM) broadcast  at 10Hz rate \cite{sae}.

\begin{figure*}
	\centering
    \includegraphics[width=170mm]{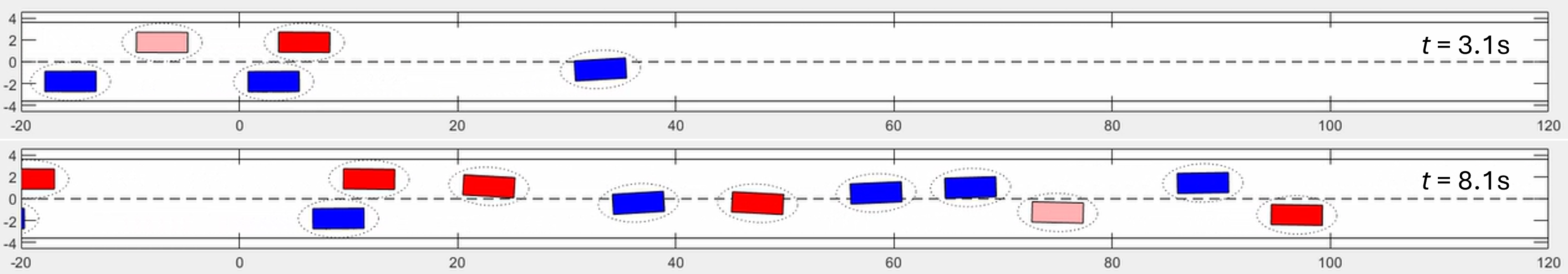}
    \vspace{-2mm}
	\caption{Two still frames taken 5s apart from an interchange traffic video. Blue vehicles go from right to left lane and and red vehicles from left to right. The pale red vehicle has lost propulsive power at -20m marker. }
	\label{fig:interchange}
    \vspace{-2mm}
\end{figure*}

The interchange traffic at two time instants separated by 5s is shown in Fig. \ref{fig:interchange} .  As the vehicles enter the interchange zone, they are commanded to change lanes while avoiding collisions or running off the road.  The pale (transparent) red vehicle is randomly selected to behave  unexpectedly (in this run, it lost acceleration capability). 

The vehicle dynamics considered in this paper is the standard simplified bicycle model \cite{rajamani}:
\begin{equation} \barr{l}{\dot x = v \cos\theta \\*[1mm]
    \dot y = v \sin\theta \\*[1mm]
    \dot \theta = \frac{v}{L_w} \delta \\*[1mm]
    \dot v = a_c} \label{bicycle} 
    \end{equation}
where $x, y$ are the (global) cartesian coordinates of the vehicle center, $v$ is its longitudinal velocity, $\theta$ is the orientation relative to the $x$ coordinate, $\delta \approx \tan(\delta)$ is the wheel steering angle, $L_w$ is the wheelbase, and $a_c$ is the vehicle acceleration.

The safety-filter setup requires a baseline controller $u_0$ that ignores obstacles and moves the vehicles along the desired path. We employ the standard pure-pursuit algorithm with independent lateral and longitudinal components. The former tracks the desired lateral position on the road with a velocity-adjusted look-ahead, while the latter regulates vehicle velocity to its desired setpoint. 

 The vehicles are protected by elliptic CBFs: the center of the other vehicle is required to stay outside of an ellipse centered on the ego vehicle (Fig. \ref{fig:interchange} shows 1/2-size ellipses  for clarity).
The CBF function $h(x)$ uses the definition of the ellipse as the set of points with the sum of distances from the two focal points equal to the length of the major axis, $2\alpha r$ ($r$ is the semi-minor axis and $\alpha$ is the ratio of the major and minor axes). Thus, the elliptic CBF for vehicles $i$ and $j$ is given by:
\begin{equation}
    h_{ij} = \|X_i +\rho \varphi_i -X_j\| + \|X_i -\rho \varphi_i -X_j\| -2\alpha r \label{h_ellipse}
\end{equation}
where $\rho = r \sqrt{\alpha^2 -1}$ is the distance of the focal points from the center of the ellipse $X_i=[x_i,y_i]^T$, and $\varphi_i = [\cos \theta_i, \ \sin \theta_i]^T$.

As shown in \cite{jankovic_ACC26}, the  control inputs appear in the second derivative of each inter-agent CBF $h_{ij}$. Combining the 
CBF and its two derivatives produces the CBF constraint: 
\begin{equation}
\ddot h_{ij} + l_1 \dot h_{ij} + l_0 h_{ij} = a_{ij} + b_{ij} u_i + b_{ji} u_j\ge 0
\label{ij-th_constraint}
\end{equation}
with $l_0$ and $l_1$ selected such that the roots of $s^2 +l_1 s +l_0 =0$ are negative real numbers. They determine the behavior of the CBF function $h_{ij}$ when the constraint is active.  

The road boundary CBFs enforce the vehicle lateral position $y$ of to satisfy $y_i \in[rb_r, rb_l]$. Thus, the pair of road  CBF constraints for the agent $i$ are 
\begin{equation}
    h_{ir} = y-rb_r \ \ {\rm and} \ \ h_{il} =- y + rb_l \label{h_road}
\end{equation}
The right road CBF constraint is obtained by combining $h_{ir}$
and its first two derivatives to obtain
\[ a_{ir} + b_{ir} u_i \ge 0 \] 
where $a_{ir} = l_1 v_i \sin\theta_i + l_0 h_{ir}, \ \  b_{ir} = [v_i^2 \cos\theta _i /L_w,\ \sin \theta_i]$ and similarly for the left road boundary.

\section{CBF Safety Filters with and without MPF}
\label{SF}

We consider three controllers (full-MPF, no-MPF, and split-MPF) with different  $w$-feedback configurations as explained below. With $\Delta = I$, i.e. without unexpected behavior, they all agree and compute the same set of inputs. 

In the decentralized PCCA \cite{santillo, jankovicTCST}, each agent independently computes everyone's control action based on the information it has and implements its own.  The full-MPF algorithm closely resembles the decentralized PCCA used in \cite{jankovic_ACC26}, except that the complete information (that is, $u_{0}$) is available: 
\begin{equation}
  \barr{l}{ \min_{u \in {\mathcal{U}}}  \sum_{i=1}^{N_a}  \|u_{i} -u_{i0}\| \ \ {\rm such \ that} \\*[2mm]
a_{ij} + b_{ij} (u_{i} + w_{i}) + b_{ji} (u_{j} + w_{j}) \ge  0 , \  i,j = 1,\ldots, N_a \\*[2mm] 
a_{i\mu} + b_{i\mu} ( u_{i} + w_{i} ) \ge  0, \ \mu =\{``r", ``l"\} }  \label{MPF}
\end{equation}
 where $N_a$ is the number of agents. The set ${\mathcal U}=\{u \in {\R}^{2N_a}: \underbar {u} \le u_i \le \bar u \} $ with $\underbar u$ and $\bar u$  the upper and lower limits on the steering and acceleration: $\pm \pi/7$rad for steering and  $[-8,\ 4]m/s^2$ 
 for acceleration, the same as in \cite{jankovic_ACC26}. In the implementation, the constraints are soft with high cost \cite{jankovic_ACC26}.

 The {\em full-MPF} determines the proprioceptive $w$-feedback inputs into the QP (\ref{MPF}) as in (\ref{w-loop}). In this case, $u^{act}$ may be determined from on-board sensors:  accelerometers, propulsion system sensors, steering angle and/or velocity measurements. 

 The pure ``centralized" approach sets $w \equiv 0$ in (\ref{MPF}) and is referred to as {\em no-MPF}. The entity that computes the control actions simply assumes they are executed faithfully.

 The previous two algorithms can be implemented as a single controller or distributed in the vehicles as exact copies. In contrast, the {\em split-MPF} needs a separate controller for each actuator, that is, it becomes decentralized not only across agents, but also across actuators with the controller (\ref{MPF}) ran for each agent-actuator combination. Compared to the centralized roadside-unit implementation, the complexity increases $2N_a$ times. Compared to the distributed approach, the complexity increases by a factor of 2.
 The size of the $w$ vector also increases from $2N_a$ to $4N_a^2$, with each agent-actuator controller having its own copy. For example, $w_{jq}^{si}$ is computed by
 \beq \epsilon \dot w_{jq}^{si} = -w_{jq}^{si} + (u_{jq}^{act} - u_{jq}^{*si}), \ \ \ w_{is}^{si} \equiv 0 \label{split-w} \eeq
where $s$ and $i$ in the superscript denote that this quantity belongs to the steering controller of the agent (vehicle) $i$ and is based on the local copy of the control action for agent $j$ and actuator $q$ (``$q$" stands for either ``$s$" (steering) or ``$a$" (acceleration)). The information needed to compute (\ref{split-w}) is contained in standard vehicle BSMs. For 16-vehicle lane-swap Monte Carlo runs reported below,  the average loop time was about 5ms, while the worst observed time was 34ms (i7-3800H, 2.5GHz laptop running Matlab {\em quadprog} solver).  The loop time corresponds to the split-MPF algorithm (2 QPs of the form (\ref{MPF})) ran on-board each vehicle. The update rate of the controller is $100$ms, consistent with the BSM rate.  The fast time constant $\epsilon$ is $200$ms.

\section{Monte Carlo Simulations}
\label{results}

 We ran Monte Carlo (MC) simulations with 16 lane-swapping vehicles, each  with  random initial positions and random initial/desired speeds in the range 20 to 24m/s. The average traffic density is 3400 veh/h per lane with Fig. \ref{fig:interchange} illustrating the resulting traffic pattern. The initial conditions were identical for runs with each of the three controllers. As in \cite{jankovic_ACC26},  $15$\%   of the vehicles, on average, are randomly selected to go straight and not change lanes. 
 
 To compare the three MPF options, we introduced three sources of unexpected behavior in the form of $\Delta$ uncertainty:
 \begin{enumerate}
     \item A single vehicle is randomly selected to have acceleration limited at or below the road load deceleration (loss of propulsion) as in Fig \ref{fig:difference}.
     \item With probability 0.5 vehicles are selected to be ``on rails" (i.e., they cannot steer to avoid other vehicles) and simultaneously have their acceleration  attenuated. 
     \item With probability 0.5 vehicles are selected to have both steering and acceleration filtered by first order filters.  
 \end{enumerate}

The MPF algorithms do not know that vehicles have their capabilities degraded.
In all three cases the resulting uncertainty $\Delta$ is passive, and in  the fist and second case it is input strictly passive. 


\subsection{Single vehicle loss of propulsion}
\label{sec:accel_clip}
\begin{figure}
\vspace{1mm}
 \hspace*{1mm}
 \includegraphics[width=80mm]{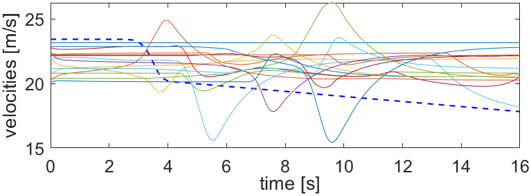}   
      \vspace{-2mm}
   \caption{Acceleration profiles for all sixteen vehicles of a single run.}
	\label{fig:clipped_accel}
   \vspace{-1mm}
\end{figure}

This is the case that illustrates the capability of MPF and, at the same time, is a realistic traffic scenario.  A single, randomly selected vehicle lost ability to accelerate 20m before the lane change command starts. Its acceleration is clipped at $-0.2m/s^2$, approximately equal to the coast-down deceleration. Steering is available as illustrated  in  Fig. \ref{fig:interchange} (the affected vehicle changes lanes) as is additional deceleration by braking, 
 Fig, \ref{fig:clipped_accel} (the affected vehicle, blue-dash trace, initially brakes more). The two figures correspond to the same simulation run. 
 
\begin{figure}
\vspace{0mm}
 \hspace*{-3mm}
 \includegraphics[width=90mm]{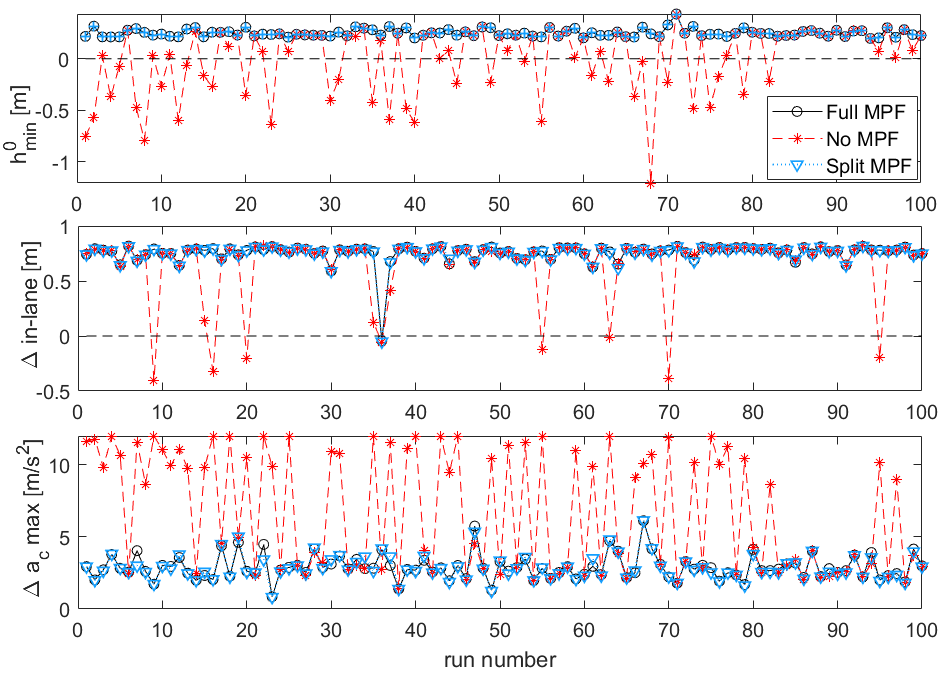}   
      \vspace{-6mm}
   \caption{Loss of propulsion. Top plot: the minimal value of $h_{ij}^0$  over all agents in a run; Middle plot: lane completion metric; Bottom plot: max $\Delta a_c$ between subsequent samples. }
	\label{fig:MC_accel_clip}
   \vspace{-2mm}
\end{figure}

The top plot in Fig. \ref{fig:MC_accel_clip} shows the values of $h^0_{min} = \min_{i,j,t} h_{ij}^0(t)$. The superscript ``0" is for the reference value for the $h_{ij}$'s. The actual CBF $h_{ij}$ used in the controller is enlarged by a  10\% barrier margin (see \cite{jankovic_ACC26}) to account for ${\mathcal O}(\epsilon)$ and 
for modeling approximations. The no-MPF algorithm has 32\% of its runs with CBF violations.
The full- and split-MPF handle this scenario without a problem thanks to their capability to react very fast to unexpected behavior.

The middle plot in Fig. \ref{fig:MC_accel_clip}  shows the lane completion metric in terms of the vehicle distance from the assigned lane boundary (the worst case in an MC run). Negative numbers mean a vehicle has not managed to get fully into the new lane by 120m marker. In this metric, no-MPF has six runs with incomplete lane changes, while full-MPF  and split-MPFs have one each. It is interesting to note that almost all vehicles manage to change lanes, even those that are impaired.

The bottom plot shows a measure of harshness of the control action.  Many of the runs with no-MPF algorithm produce full acceleration reversals (from full acceleration to full deceleration or vice-versa, $\Delta a_c = 12m/s^2$) in one sample (100ms).   The  full- and split-MPFs are much milder. Their traces in Fig. \ref{fig:MC_accel_clip} are almost indistinguishable. 

The statistics for 100 MC runs is shown in Table \ref{tab:accel_clip}. The items in {\bf bold} font indicate undesirable outcomes. The first column identifies the algorithm. The second column corresponds to the minimal value in the top plot in Fig. \ref{fig:MC_accel_clip}.  
As a  rule of thumb, $h_0 = 0$ means that the ellipses in Fig. \ref{fig:interchange} touch.  The value $h^0_{\min} = 0.20$m is the same as for the nominal,  no-impairment case.  The third column shows the number of incomplete lane changes. The forth, ``OOB" column shows the worst case distance a vehicle went off the road. The column five corresponds to the worst value of $\Delta a_c$ from the bottom plot in Fig. \ref{fig:MC_accel_clip}. The last column shows   the average number of times per run that $\Delta a_c$ exceeded $2m/s^2$. 


\begin{table}
	\caption{100 MC runs with loss of propulsion for a random vehicle}	
    \vspace{-0.1in}
    \hspace{1mm}
	\begin{tabular}{|l|c|c|c|c|c|}
	\cline{1-6}
	                 Algorithm                                   & min $h_0$ & incomp- & OOB & max $\Delta a_c$\! &  $\Delta a_c > 2$\!  \\ 
                      &   $[m]$      &          lete  LS    \#  & $[m]$  & $[m/s^2]$  & \#  \\ \hline
	\multicolumn{1}{|l|}{full MPF}       & \multicolumn{1}{c|}{0.20}  & {\bf 1} & N/A    & 6.1        & 22          \\ \hline
	\multicolumn{1}{|l|}{no MFP}            & \multicolumn{1}{c|}{\bf --1.21}  & {\bf 6}  & {\bf 2.0}      & 12          & 52        \\ \hline
    \multicolumn{1}{|l|}{split MFP}          & \multicolumn{1}{c|}{0.20} & {\bf 1}  & N/A      & 6.2          & 20        \\ \hline
	\end{tabular}
	\vspace{-3mm}
	\label{tab:accel_clip}
\end{table}


\subsection{Vehicles on rails with attenuated acceleration}
\label{sec:on_rails}
\begin{figure}
\vspace{-0mm}
 \hspace*{-3mm}
 \includegraphics[width=90mm]{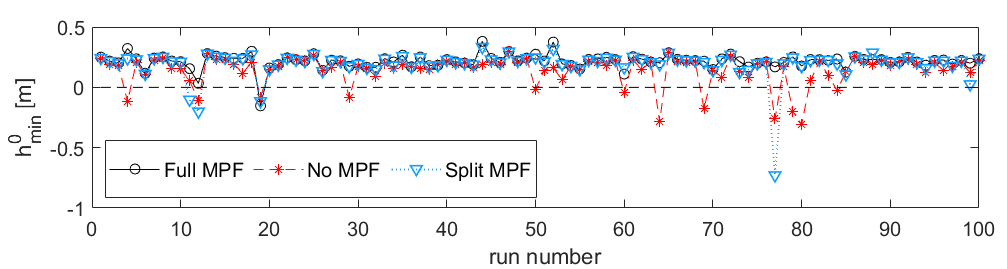}  
      \vspace{-6mm}
   \caption{ Random vehicles on-rails and with attenuated acceleration: Minimal value of $h_{ij}^0$ barrier over all agents in a run.}
	\label{fig:MC_on_rail}
   \vspace{-3mm}
\end{figure}

In contrast to the loss-of-propulsion case, we consider a less realistic scenario selected as unfavorable for the split-MPF algorithm. Vehicles are randomly selected with probability 0.5 to be on rails (i.e., selected vehicles use  $u_{0}$ instead of $u^*$ steering) and at the same time have the acceleration and braking reduced proportionaly: $\Delta_{i2} = 0.7$. Hence, if two adjacent vehicles are impaired, all four actuators engaged in collision avoidance are impacted, while the resulting $\bar \Delta$ is input strictly passive. 
Full-MPF acceleration is controlled by a system that knows if agent's acceleration is impaired or not, while ego split-MPF acceleration controller does not. As a result, split-MPF does have four runs with $h_0 < 0$ shown in Fig. \ref{fig:MC_on_rail}. The statistics is summarized in Table \ref{tab:on_rails}.
The severe impairment presents a problem for all algorithms, but full-MPF handles it with the smallest impact.

\begin{table}
\vspace*{-3mm}
	\caption{100 MC runs: vehicles on rails with attenuated acceleration}	
     \vspace{-2mm}
     \hspace{1mm}
	\begin{tabular}{|l|c|c|c|c|c|}
	\cline{1-6}
	                 Algorithm                                   & min $h_0$ & incomp- & OOB & max $\Delta a_c$\! &  $\Delta a_c > 2$\!  \\ 
                      &   $[m]$      &          lete  LS    \#  & $[m]$  & $[m/s^2]$  & \#  \\ \hline
	\multicolumn{1}{|l|}{full MPF}       & \multicolumn{1}{c|}{\bf -0.15}  & {0} & {\bf 0.1}    & 8.9        & 33          \\ \hline
	\multicolumn{1}{|l|}{no MFP}            & \multicolumn{1}{c|}{\bf --0.31}  & 0  & N/A     & 5.8          & 20        \\ \hline
    \multicolumn{1}{|l|}{split MFP}          & \multicolumn{1}{c|}{\bf -0.73} & {\bf 2}  & N/A      & 11.7          & 25        \\ \hline
	\end{tabular}
	\vspace{-3mm}
	\label{tab:on_rails}
\end{table}

\subsection{Filtered steering and acceleration}
\label{sec:filtered_inputs}

\begin{figure}
\vspace{2mm}
 \hspace*{-2mm}
 \includegraphics[width=90mm]{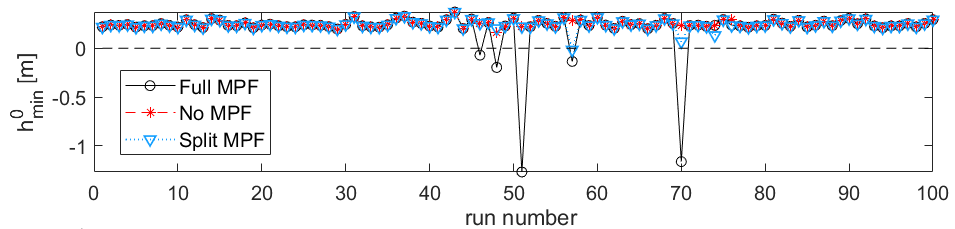}   
      \vspace{-6mm}
   \caption{Filtered steering and acceleration: minimal value of $h^0_{ij}$ barrier over all agents in a run.}
	\label{fig:MC_filters}
   \vspace{-3mm}
\end{figure}

In this case $\Delta_{i1}(s) = \frac{1}{0.2s+1}$ and $\Delta_{i2} (s)= \frac{1}{0.4s+1}$ are filter transfer functions for steering and acceleration for randomly selected vehicles (0.5 probability). 
The results of 100 MC runs are shown in Fig. \ref{fig:MC_filters} and Table \ref{tab:filters}. In this case the full-MPF runs into safety violations because the absence of strict input passivity leads to oscillatory behavior (the fast $w$-loop response is stable, but under-damped). Split-MPF fares much better. The no-MPF has no issues with the actuator filters that are much faster than  the dominant time constant $\tau_h = 2.5s$ of the inter-agent CBF constraint. 

\begin{table}
	\caption{100 MC runs with random filtered steering and acceleration}	
    \vspace{-1mm}
    \hspace{1mm}
	\begin{tabular}{|l|c|c|c|c|c|}
	\cline{1-6}
	                 Method  and                                   & min $h_0$ & incomp- & OOB & max $\Delta a_c$\! &  $\Delta a_c > 2$\!  \\ 
                    use case  &   $[m]$      &          lete  LS    \#  & $[m]$  & $[m/s^2]$  & \#  \\ \hline
	\multicolumn{1}{|l|}{Full MPF}       & \multicolumn{1}{c|}{\bf --1.27}  & { 0} & {\bf 5.4}    & 11.4        & 42          \\ \hline
	\multicolumn{1}{|l|}{No MPF}            & \multicolumn{1}{c|}{0.16}  & {\bf 1}  & {N/A}      & 6.8          & 16        \\ \hline
    \multicolumn{1}{|l|}{Split MPF}          & \multicolumn{1}{c|}{\bf --0.02} & {\bf 1}  & N/A      & 11          & 34        \\ \hline
	\end{tabular}
	\vspace{-3mm}
	\label{tab:filters}
\end{table}

Interesting points about the interchange traffic simulations:
\begin{itemize}
    \item The analytical results of Sections \ref{sec:singular_p} and \ref{sec:split_MPF} predict well the results of the interchange lane swap simulations. 
    \item The speed reduction needed to execute the lane swap is minimal: on average, $-0.6$ mi/h (49.5 to 48.9 mph) for the loss-of-propulsion  and $-0.1$ mi/h for the other two cases with no difference between the  algorithms.
    \item A case could be made that the split-MPF provides the strongest resilience to unexpected events and robustness to unmodeled dynamics. It should be emphasized that it can only be implemented as a {\em decentralized} controller.
    \item Some of the benefits of MPF may be attributed to the use of the acceleration signal not employed by the no-MPF algorithm. It appears that human drivers leverage acceleration (vestibular sense) to improve performance, see \cite{markkula} and reference therein.
    \item Vestibular sense in humans does not replace proprioception. Likewise, adding acceleration as a measured state to a  CBF constraint would not provide the MPF functionality in terms of speed of response and resilience. 
\end{itemize}

\section{Conclusion}
 In this paper we present a machine proprioceptive feedback mechanism for motion control intended for managing sudden unexpected events. In many respects, MPF resembles human proprioception -- our fast-reacting safety system for gait and position.  The challenge of  assuring stability of the fast system is addressed with a biology-inspired split-MPF algorithm that is, by necessity, decentralized. In highway interchange traffic with randomly impaired vehicles, the two MPF versions showed safer and more resilient behavior  compared to the no-MPF baseline.  


\section{Acknowledgment}

The author would like to thank Dr. Ivana Jankovic, MD, for informative discussion and useful pointers on human proprioception.


\bibliographystyle{IEEEtran}
\bibliography{mybibfile}

\end{document}